\documentclass[useAMS,usenatbib]{mn2e}
\pdfoutput=1

\usepackage{times}
\usepackage{amsmath, amssymb}
\usepackage{graphicx}
\usepackage{hyperref}
\usepackage{float}
\usepackage[caption=false]{subfig}

\providecommand{\eprint}[1]{\href{http://arxiv.org/abs/#1}{#1}}

\newcommand{\ffp}{{FFP6 }}
\newcommand{\sevem}{SEVEM }
\newcommand{\mask}{U73 }

\newcommand{\maketitlenonewpage}{{\let\newpage\relax\maketitle}}
\newcommand{\mlv}{\textbf{m}_{\rm r}}
\newcommand{\mmlv}{\bar{\textbf{m}}_{\rm r}}
\newcommand{\deltaTT}{\frac{\Delta T}{T}}
\newcommand{\deltaTTv}{\left.\deltaTT\right\vert}
\newcommand{\rdisk}{r_{\rm disk}}
\newcommand{\Alv}{A_{\rm LV}}
\newcommand{\AT}{A_{\rm T}}

\def\apj{ApJ}

\def\apjs{ApJS}

\def\prd{Phys.~Rev.~D}

\def\jcap{Journal of Cosmology and Astroparticle Physics}

\title[Local variance asymmetries in Planck temperature anisotropy maps]{Local variance asymmetries in Planck temperature anisotropy maps}
\author[Saroj Adhikari]{Saroj Adhikari$^{1}$\thanks{E-mail:
sza5154@psu.edu}\\$^{1}$Institute for Gravitation and the Cosmos, The Pennsylvania State University, University Park, PA 16802, USA}

\begin{document}
\maketitle
\label{firstpage}

\begin{abstract}
 Recently, it was shown that local variance maps of temperature anisotropy are simple and useful tools for the study of large scale hemispherical power asymmetry. This was done by studying the distribution of dipoles of the local variance maps. In this work, we extend the study of the dipolar asymmetry in local variance maps using foreground cleaned Planck 143 GHz and 217 GHz data to smaller scales. In doing so, we include the effect of the CMB Doppler dipole. Further, we show that it is possible to use local variance maps to measure the Doppler dipole in these Planck channel maps, after removing large scale features (up to $l=600$), at a significance of about $3 \sigma$. At these small scales, we do not find any power asymmetry in the direction of the anomalous large scale power asymmetry beyond that expected from cosmic variance. At large scales, we verify previous results i.e. the presence of hemispherical power asymmetry at a significance of at least $3.3 \sigma$. 
\end{abstract}


\section{Introduction}
A number of studies have been performed that show approximately 3.5$\sigma$ hemispherical power asymmetry at large scales in WMAP and Planck CMB temperature fluctuations \cite[]{2004ApJ...605...14E, 2007ApJ...660L..81E, PlanckCollaboration2013, 2013JCAP...09..033F, 2009ApJ...699..985H}. Recently, \cite{2014ApJ...784L..42A} used a conceptually simple pixel space \textit{local variance} method to demonstrate the presence of asymmetry at large scales in WMAP and Planck data at a significance of at least $3.3 \sigma$; they find that none of the 1000 isotropic Planck \ffp simulations had a local variance dipole amplitude equal to or greater than that found in data for disk radii $6^\circ \leq \rdisk \leq 12^\circ$. In this method, a local variance map ($\mlv$) is generated at a smaller HEALPIX \cite[]{healpix} resolution $N_{\rm side}$ from a higher resolution CMB temperature fluctuations map by computing temperature variance inside disks of certain radius $\rdisk$ centered at the center of each pixel of the HEALPIX map with resolution $N_{\rm side}$. Isotropic simulations are used to get the expected mean map ($\mmlv$), and each map is normalized:\begin{eqnarray}\mlv^{n}&=&\frac{\mlv-\mmlv}{\mmlv}.\end{eqnarray} Then, we obtain local variance dipoles by fitting for dipoles in each of these normalized maps (both simulations and Planck data) using the HEALPIX \texttt{remove\_dipole} module with inverse variance (of the 1000 simulated local variance maps) weighting. In this paper, we will denote the amplitude of a local variance dipole as $\Alv$. The local variance dipole obtained from data is then compared to the distributions obtained from simulations. 

In this work, we make use of the local variance method and extend the results obtained in \cite{2014ApJ...784L..42A} to include smaller disk radii. The authors in \cite{2014ApJ...784L..42A} focused on large scale power asymmetry, and therefore only looked at large values of $\rdisk$. After confirming their results at large disk radii ($\rdisk \geq 4^\circ$), we perform the same local variance dipole analysis using smaller disk sizes. We find that, for smaller disk radii, the contribution of the \textit{Doppler dipole} becomes increasingly significant. The Doppler dipole in the local variance map is an expected signal because of our velocity with respect to the CMB rest frame. While the direction and magnitude of the CMB dipole has been known from previous CMB experiments \cite[]{2011ApJS..192...14J}, the Doppler dipole signal in the temperature fluctuations is rather weak and reported only recently by the Planck Collaboration \cite[]{2013arXiv1303.5087P} using harmonic space estimators \cite[]{2011JCAP...07..027A, 2011PhRvL.106s1301K}. We work towards detecting the expected Doppler dipole in the local variance maps after removing large scale features from the maps. Our goal therefore is two fold: first, extend the local variance dipole study of the hemispherical power asymmetry to smaller disk radii and second, use the method of local variance to detect the Doppler dipole whose amplitude is much smaller but is expected to contribute at all angular scales.

Before presenting the details of the analysis and results, we would like to point out and clarify a difference between our analysis and that of Akrami et al. They used 3072 disks ($N_{\rm side}=16$ healpix map) for all sizes of disks they considered ($\rdisk \geq1^\circ$). However, we find that for $\rdisk =1^\circ$ and $2^\circ$, 3072 disks are not enough to cover the whole sky. Therefore, we use  $N_{\rm side}=32$ (12288 disks) for $\rdisk=2^\circ$ and $N_{\rm side}=64$ (49152 disks) for $\rdisk=1^\circ$. Once we do this, we find that, unlike the results in Akrami et al. (see Fig 2(a) and Table 1 in \cite{2014ApJ...784L..42A}), none of our 1000 isotropic simulations produce a local variance dipole amplitude larger than that of our foreground cleaned channel maps, for $\rdisk=1, 2^\circ$.  In fact, the effect of the anomalous dipole (with respect to the isotropic case) can be observed for even smaller angular disk radii (see Figure \ref{fig:subdegrees}). This result, however, is not surprising because the local variance dipoles computed at smaller disk radii get some contribution from the large scale anomalous power asymmetry at low $l$.

Next, in section \ref{sec:sim} we discuss in some detail the Planck CMB data and simulations used in this paper. Then, in section \ref{sec:results} we present our results for both the anomalous dipole and the Doppler dipole, followed by discussions and a summary of our results in section \ref{sec:discussion}.

\section{Simulations and data}
\label{sec:sim}
For our analysis, we use 1000 realizations of \ffp  Planck simulations \footnote{\url{http://crd.lbl.gov/groups-depts/computational-cosmology-center/c3-research/cosmic-microwave-background/cmb-data-at-nersc/}} that include lensing and instrumental effects. The \ffp simulations (CMB and noise realizations separately) are provided at each Planck frequency. We generate local variance maps from which we obtain local variance dipole distributions following the method briefly introduced in the previous section (described in more detail in \cite{2014ApJ...784L..42A}). Since the \ffp simulations for CMB and noise processed through the component separation procedure are not yet publicly available, we will generate foreground cleaned CMB channel maps at two frequencies 143 GHz and 217 GHz using the \sevem method \cite[]{2013arXiv1303.5072P, Kim2013}. We use the same four template maps as the Planck Collaboration; these are difference maps of two near frequency channels: (30-44) GHz, (44-70) GHz, (545-353) GHz and (857-545) GHz. Before generating the four templates, we smooth the larger frequency channel map to the resolution of the smaller frequency channel map: $a_{lm}^{\rm large}\rightarrow a_{lm}^{\rm large}\times B_{l}^{\rm small}/B_{l}^{\rm large}$, where $B_l$ is the beam transfer function for the given frequency channel map. See Appendix C of \cite{2013arXiv1303.5072P} for details of this method. To summarize: the cleaned 143 GHz or 217 GHz map $d_{\nu}^{\rm clean}$, is obtained by subtracting from the uncleaned channel map $d_\nu$, a linear combination of the templates $t_j$:
\begin{eqnarray}
 d_{\nu}^{\rm clean} &=& d_\nu - \sum_{j=1}^{n_t} \alpha_j t_j
\end{eqnarray}

The coefficients $\alpha_j$'s are obtained by minimizing the variance of $d_{\nu}^{\rm clean}$ outside of the mask used for our analysis. Once the linear coefficients for the four template maps are obtained, we use the same coefficients to process combinations of noise \ffp simulations in order to generate noise maps for our foreground cleaned simulated maps. We process the Planck maps and the simulated maps identically in each step. In addition to the isotropic local variance maps (obtained from isotropic realizations of CMB plus noise), we obtain two new sets of simulated local variance maps from two other models which are obtained from the isotropic realizations as explained below:
 \begin{itemize}\item{\textbf{ Doppler model}: The expected dipolar temperature modulation from the Doppler effect, with an amplitude $0.00123 b_\nu$ along the direction $(l,b)=(264^\circ, 48^\circ)$ \cite[]{2011ApJS..192...14J}. We take $b_\nu=1.96$ for the 143 GHz map and $b_\nu=3.07$ for the 217 GHz map \cite[]{2013arXiv1303.5087P}. The dipolar modulation is generated in pixel space simply as:
 \begin{eqnarray}
  \deltaTTv_{\rm dop}(\hat{n})&=&\left[1+1.23\times 10^{-3} b_{\nu} (\hat{n}\cdot\hat{p})\right] \deltaTTv_{\rm iso} (\hat{n})
 \end{eqnarray}
\begin{figure}
\centering
 \includegraphics[scale=0.55]{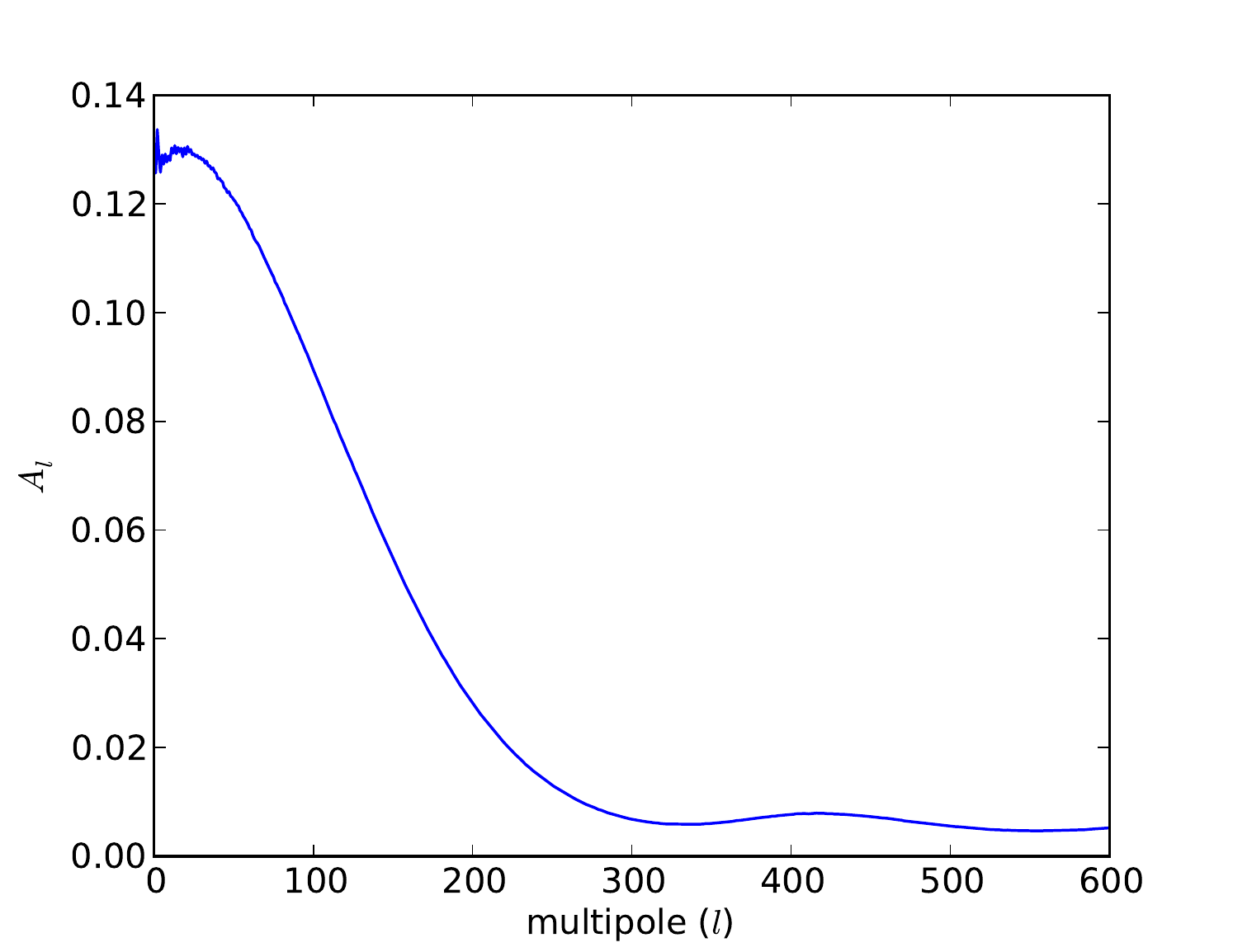}
 \caption{The hemispherical asymmetry in $C_l$s generated by our modulation model that modulates large scales (1.25$^\circ$ smoothing) with an amplitude $A_T=0.073$ in real space, in addition to the dipolar modulation at all scales due to the CMB Doppler effect.  This plot was generated using 1000 simulations for the 217 GHz channel.}
 \label{fig:modulation}
\end{figure}
 \item \textbf{Modulation model}}: The Doppler model as described above plus a large angle modulation, corresponding to a modulation at smoothing ${\rm fwhm}=1.25^\circ$, along the direction \texttt{$(l,b)=(218^\circ, -20^\circ)$}, with an amplitude $A_T=0.073$ \cite[]{PlanckCollaboration2013}. The map obtained is:
 \begin{eqnarray}
  \deltaTTv_{\rm mod}(\hat{n}) &=& \deltaTTv_{\rm dop} (\hat{n}) + A_T (\hat{n}\cdot\hat{p})\left.\deltaTT\right\vert_{\rm dop}^{\rm smoothed}(\hat{n}) \nonumber\\
  \label{eqn:modulation}
 \end{eqnarray}
 in which the modulation is only applied to the Doppler model after smoothing at ${\rm fwhm}=1.25^\circ$. Here we have used the notation $\AT$ for the amplitude of modulation for temperature fluctuations in the above equation to distinguish it from the dipole modulation amplitude in local variance maps $\Alv$. Our choice of the smoothing scale ${\rm fwhm}=1.25^\circ$ is guided by our attempt to fit the obtained local variance asymmetry in data for different $\rdisk$ that we have considered. For example, we find that a larger ${\rm fwhm}=5^\circ$ modulation does not reproduce the local variance asymmetry in data for $\rdisk=1^\circ$ or smaller. Similarly, using a smaller fwhm produces local variance asymmetry distributions with amplitudes too large to be consistent within $2\sigma$ with that seen in data for some disk radii ($\rdisk=1^\circ\; {\rm and\;} 4^\circ$, for example). Note that when we apply modulation at large scales in this manner, we are using a filter that will suppress modulation at scales much smaller than the smoothing fwhm specified. Therefore, by construction, the modulation is only generated at large scales. In Figure \ref{fig:modulation}, we plot the asymmetry generated by our modulation model in harmonic space (up to $l=600$); the quantity plotted is:
 \begin{eqnarray}
 	A_l = 2\frac{C_l^{+} - C_l^{-}}{C_l^{+}+C_l^{-}}
 \end{eqnarray}
 where $+$ is taken to be the hemisphere centered at $(l,b)=(218^\circ,-20^\circ)$ and $-$ is the opposite hemisphere. The $C_l$s for the relevant hemisphere is computed by masking the rest of the sky.
\end{itemize}

\begin{figure*}
 \centering
 \subfloat[143 GHz]{\includegraphics[scale=0.33]{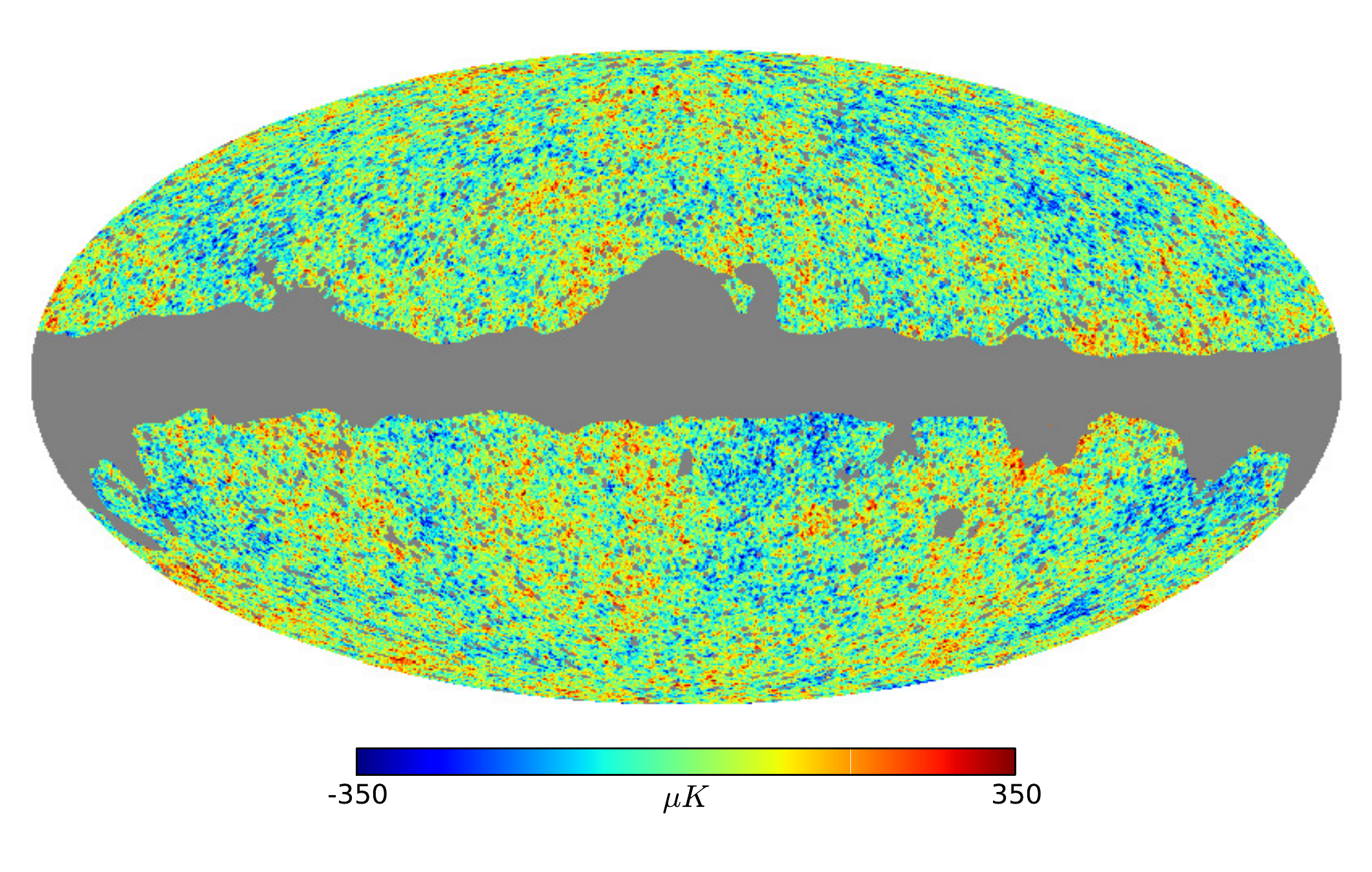}}
 \subfloat[217 GHz]{\includegraphics[scale=0.33]{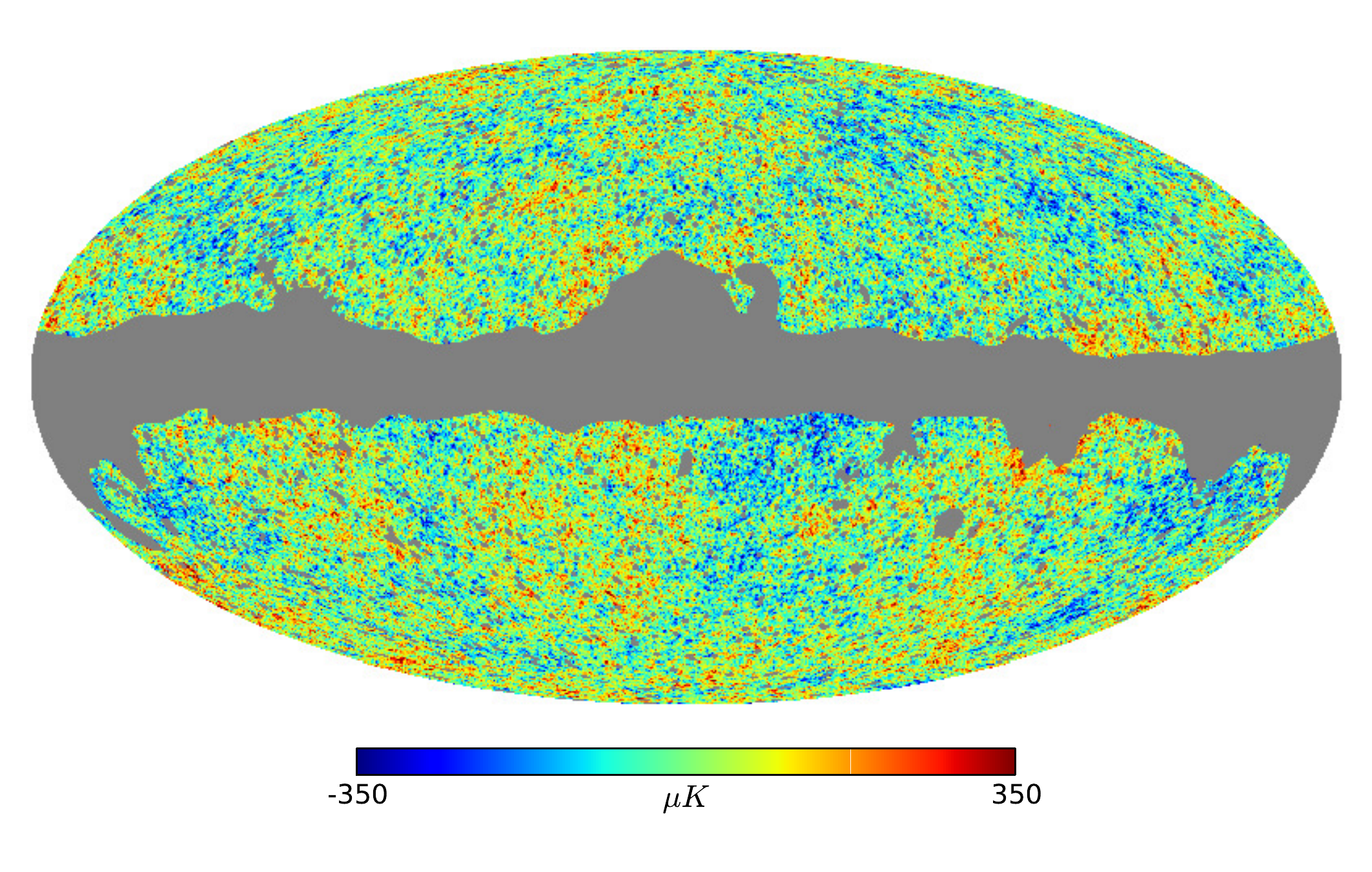}} 
 \caption{Foreground cleaned maps at channels 143 GHz and 217 GHz. Planck \mask mask is used in both maps.}
 \label{fig:cleanmaps}
\end{figure*}

\paragraph*{Masks:} For each map, we use the union mask \mask from the Planck Collaboration. This is the union of the qualitative masks for the four different component separation techniques employed by the Planck Collaboration, and leaves approximately 73 percent of sky unmasked. The template fitting in our SEVEM-like cleaning procedure is also done in the region outside this mask. The resulting clean maps at 143 GHz and 217 GHz (with the \mask mask) are shown in Figure \ref{fig:cleanmaps}.

\begin{figure*}
 \centering
  \includegraphics[scale=0.37]{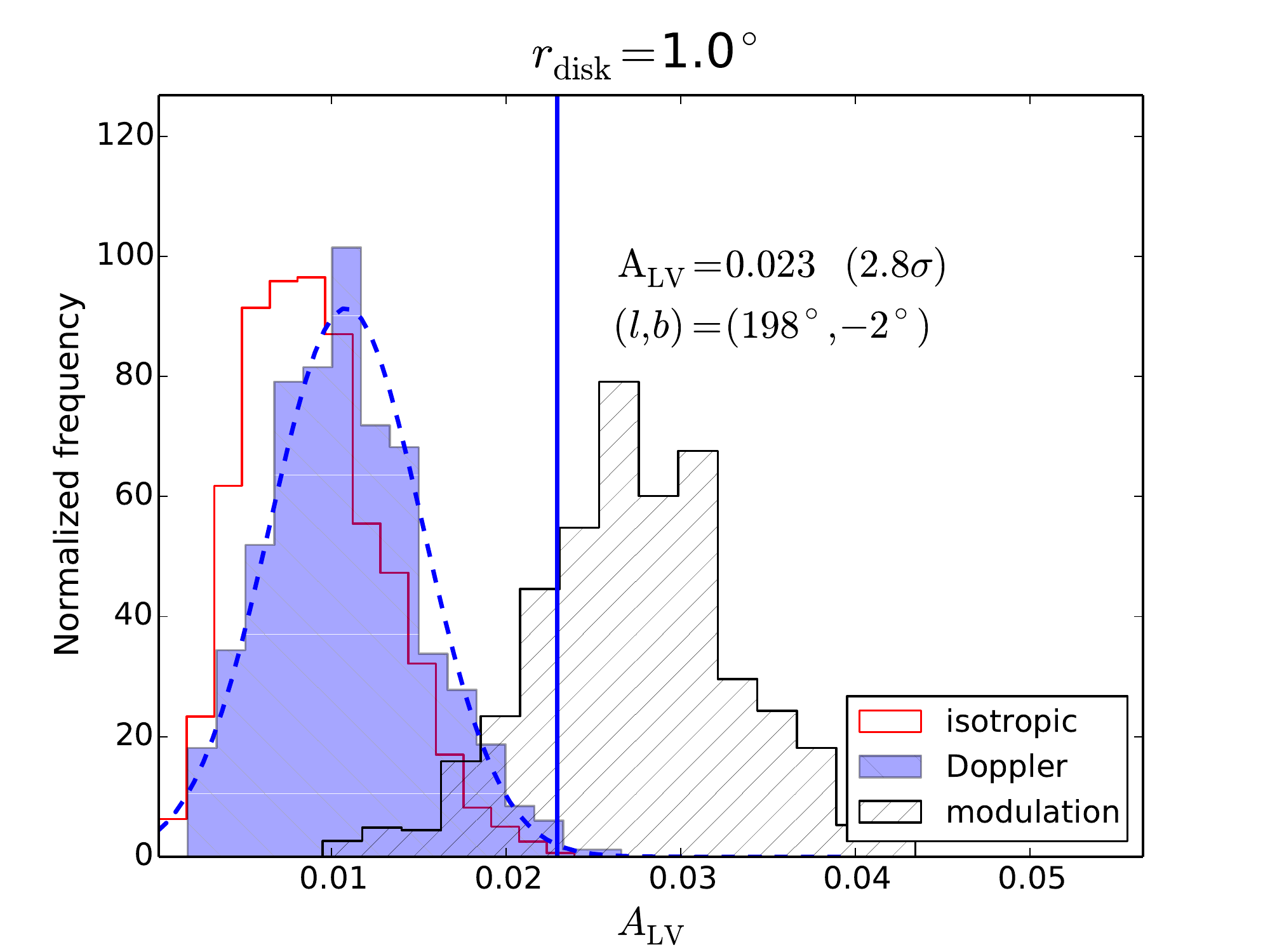}
  \includegraphics[scale=0.37]{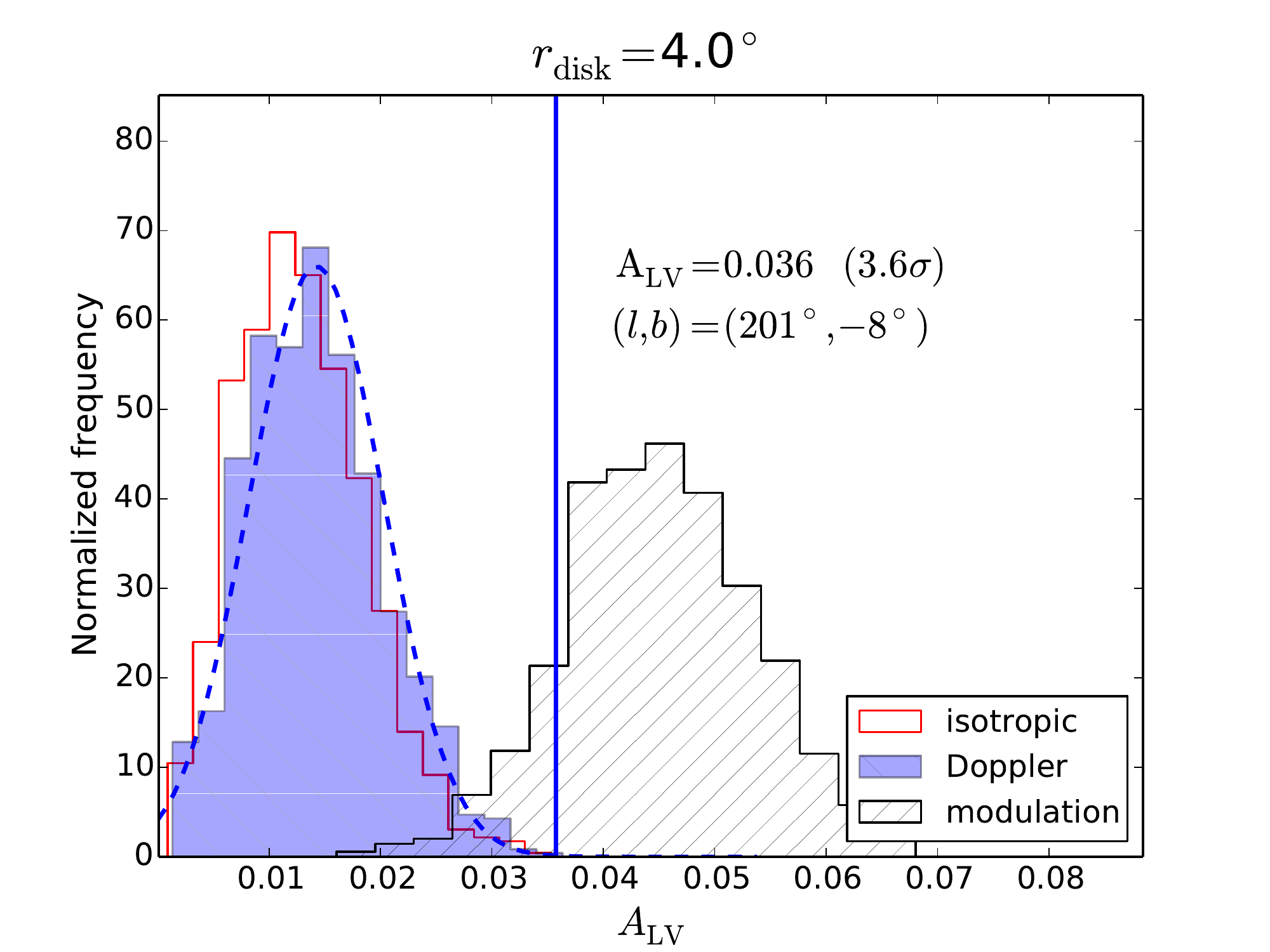}
 \caption{Local variance dipole analysis for $\rdisk=1,4$ degrees. The vertical line and the text describe the measurements from Planck maps. The histograms are obtained from the simulations for three cases---isotropic, Doppler and large scale modulation models---as labeled. For $\rdisk=2^\circ$ (not shown), we get local variance dipole of amplitude $\Alv=0.026\;(2.8\sigma)$ in the direction $(l,b)=(203^\circ,-1^\circ)$. The significance values in the above figures are computed using the Doppler model distributions. If instead we use the isotropic distributions, we obtain values of $3.5\sigma$, $3.4\sigma$ and $4.0\sigma$ for $\rdisk=1^\circ$, $2^\circ$ and $4^\circ$ respectively.}
 \label{fig:onetwodegrees}
\end{figure*}

\section{Results}
\label{sec:results}
First, let us comment that we get results consistent with \cite{2014ApJ...784L..42A} for $\rdisk=(4^\circ$ to $16^\circ)$, in terms of the significance of the magnitude, and the direction of the anomalous dipole. Also, our modified set of simulations that includes temperature modulation at large angular scales produces dipole distributions consistent with data. All our results presented in this paper were analyzed using the foreground cleaned 217 GHz maps unless otherwise stated. For large scale results i.e. relating to the anomalous dipole, we get similar results using the 143 GHz maps.

\subsection{Local variance dipole for $\rdisk=1^\circ, 2^\circ$}
For $\rdisk=1^\circ, 2^\circ$, our results in Figure \ref{fig:onetwodegrees} indicate that the effect of the Doppler dipole is increasing compared to larger disk radii. Further, the results also show that the modulation model that we have chosen (modulated at large scales corresponding to a smoothing of ${\rm fwhm}=1.25^\circ$, with an amplitude $A_T=0.073$) is consistent with data, while the no modulation (isotropic) model is disfavored at approximately $3.5 \sigma$. Note that the distribution of $\Alv$ is not exactly Gaussian, but we assume the distribution to be Gaussian to obtain standard deviation significance values as rough guides. 

Next, we extend the analysis to apply to even smaller disk radii.
\begin{figure*}
 \centering
  \includegraphics[scale=0.37]{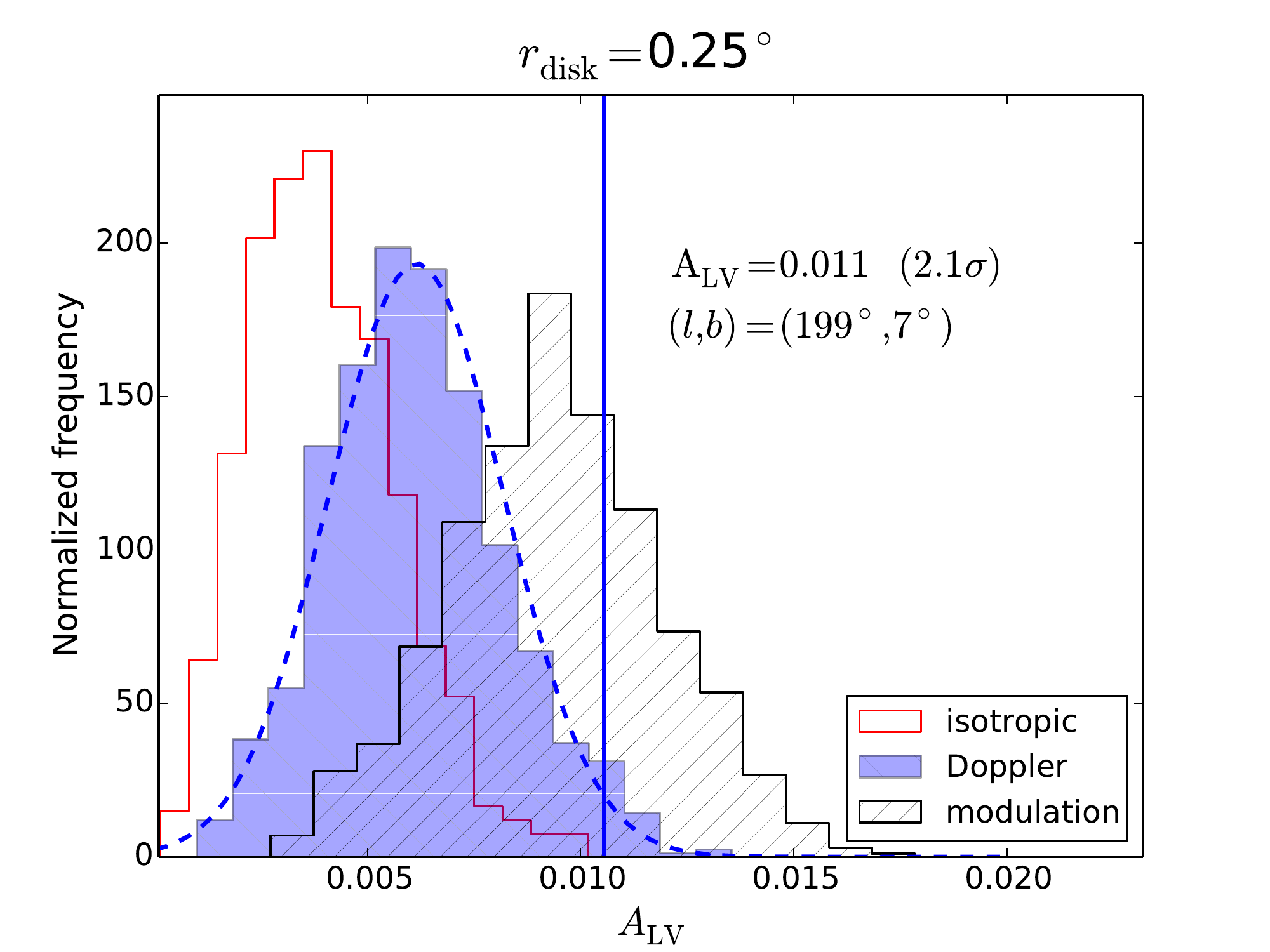}
  \includegraphics[scale=0.37]{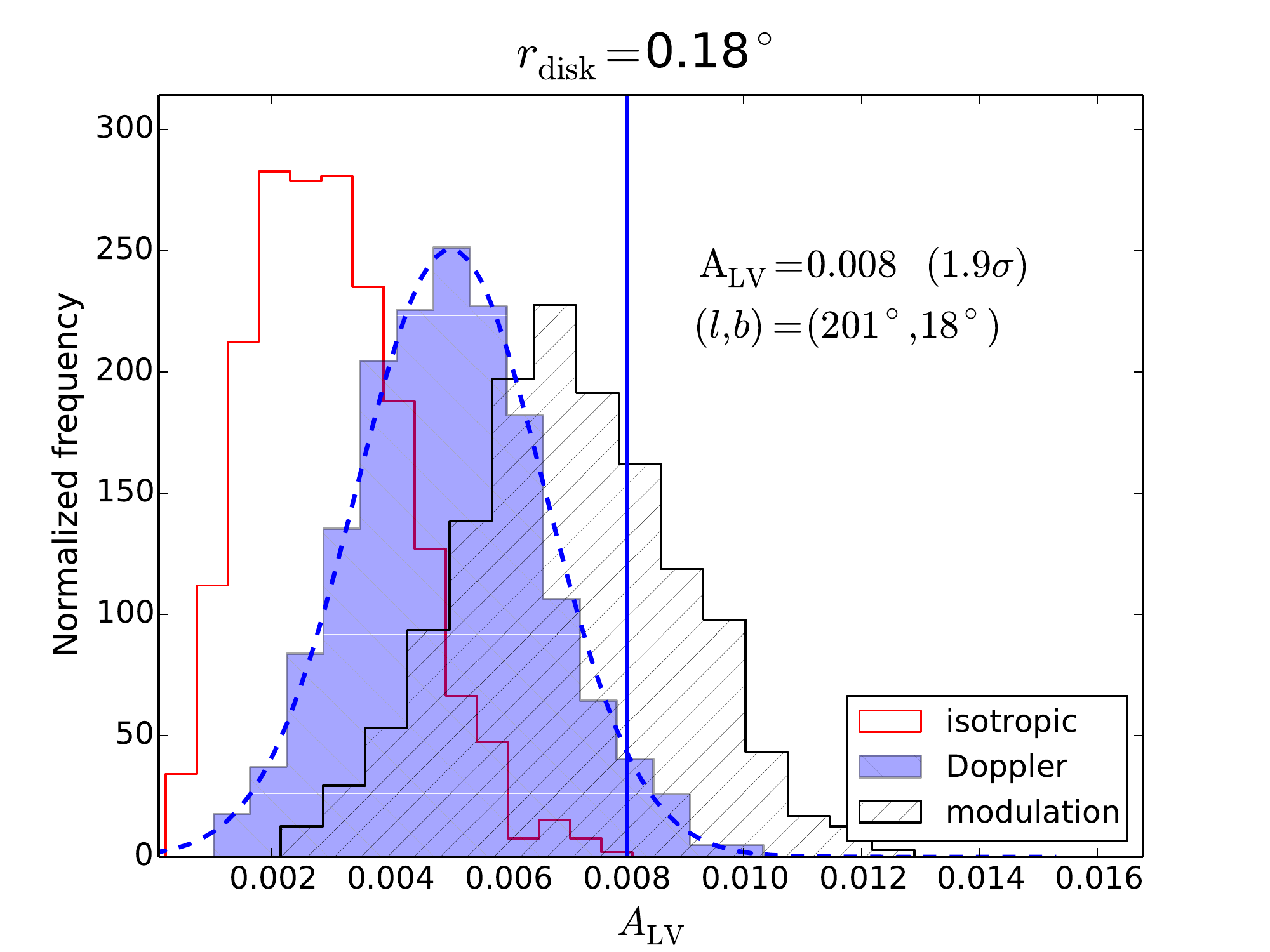}
 \caption{Local variance dipole analysis for $\rdisk=0.25,0.18$ degrees. The vertical line and the text describe the measurements from Planck maps. The histograms are obtained from the simulations for our three different models as labeled. The significance values in the figures are calculated using the Doppler model distributions. If instead we use the isotropic distributions, the corresponding significances become $3.78\sigma$ and $3.82\sigma$ respectively.}
 \label{fig:subdegrees}
\end{figure*}

\subsection{Local variance dipole for sub-degree disk radii}
We will now consider disks of radii $\rdisk=0.25^\circ$ and $0.18^\circ$. This requires that we increase the number of disks to cover the whole sky. We therefore use disks centered on pixels of a healpix map of $N_{\rm side}=256$ (786432 disks). Figure \ref{fig:subdegrees} shows our results for these two disk sizes. From the figures, it is clear that for these disk radii, the effect of the Doppler dipole cannot be neglected when computing the significance of the dipole amplitude obtained in the data. If we do not include the Doppler effect, then the local variance dipole in the Planck map is anomalous at approximately $3.8 \sigma$ for both $\rdisk=0.25^\circ$ and $0.18^\circ$, whereas with respect to the Doppler model distribution the significance drops to about $2\sigma$. Also, the direction of the dipole detected from Planck maps gets closer to the Doppler dipole as the disk radius is decreased. This is illustrated in Figure \ref{fig:dirs}. However, as seen in Figure \ref{fig:subdegrees}, the effect is still small and the distributions of the isotropic simulations and the Doppler model simulations overlap quite a bit. We will now investigate if, by removing large scale features from the temperature fluctuation maps, it is possible to separate the Doppler modulated local variance dipole distribution from the isotropic one, and therefore measure the Doppler dipole in the Planck maps.

\begin{figure*}
 \centering
 \includegraphics[scale=0.37]{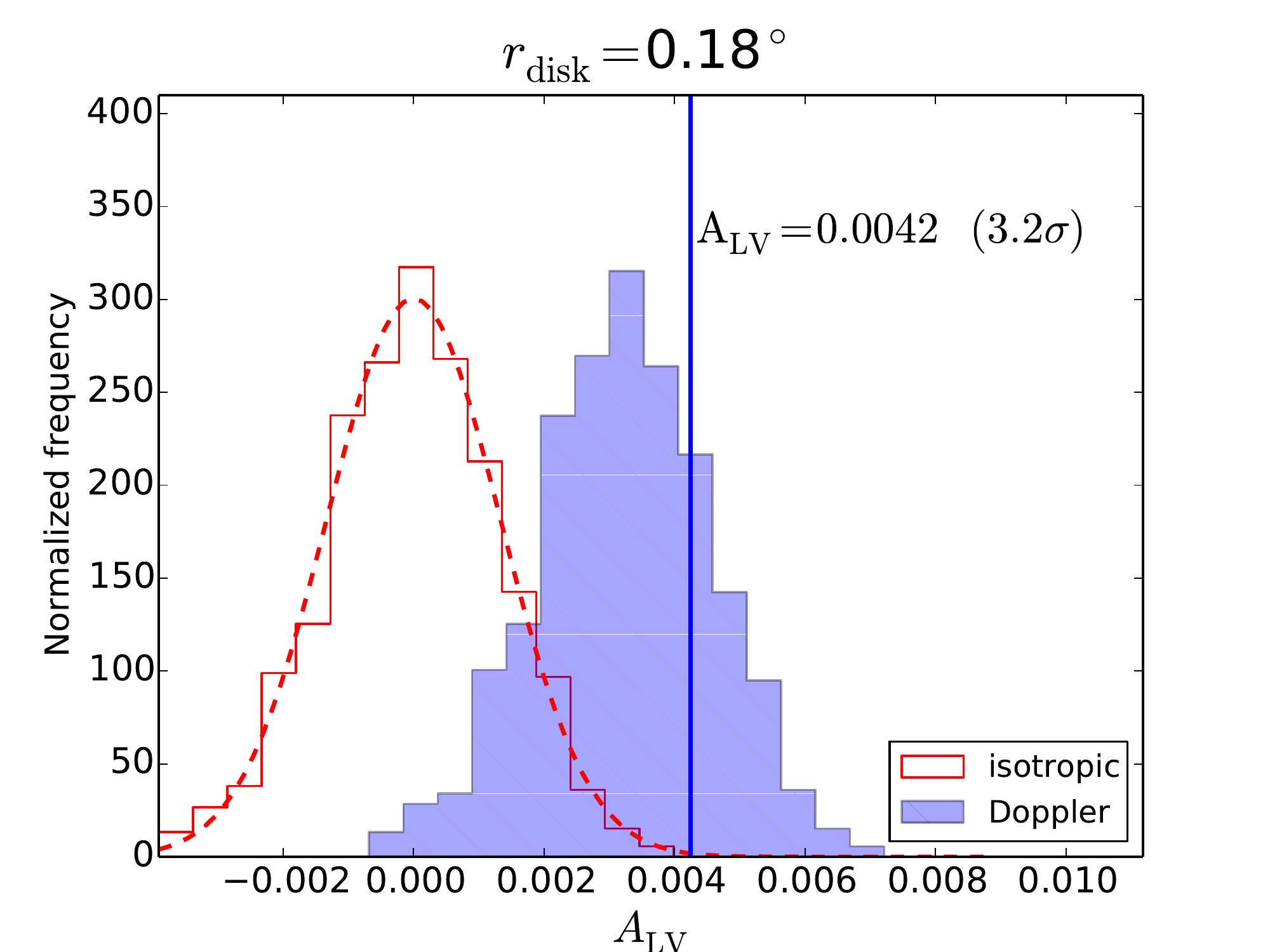}
 \includegraphics[scale=0.37]{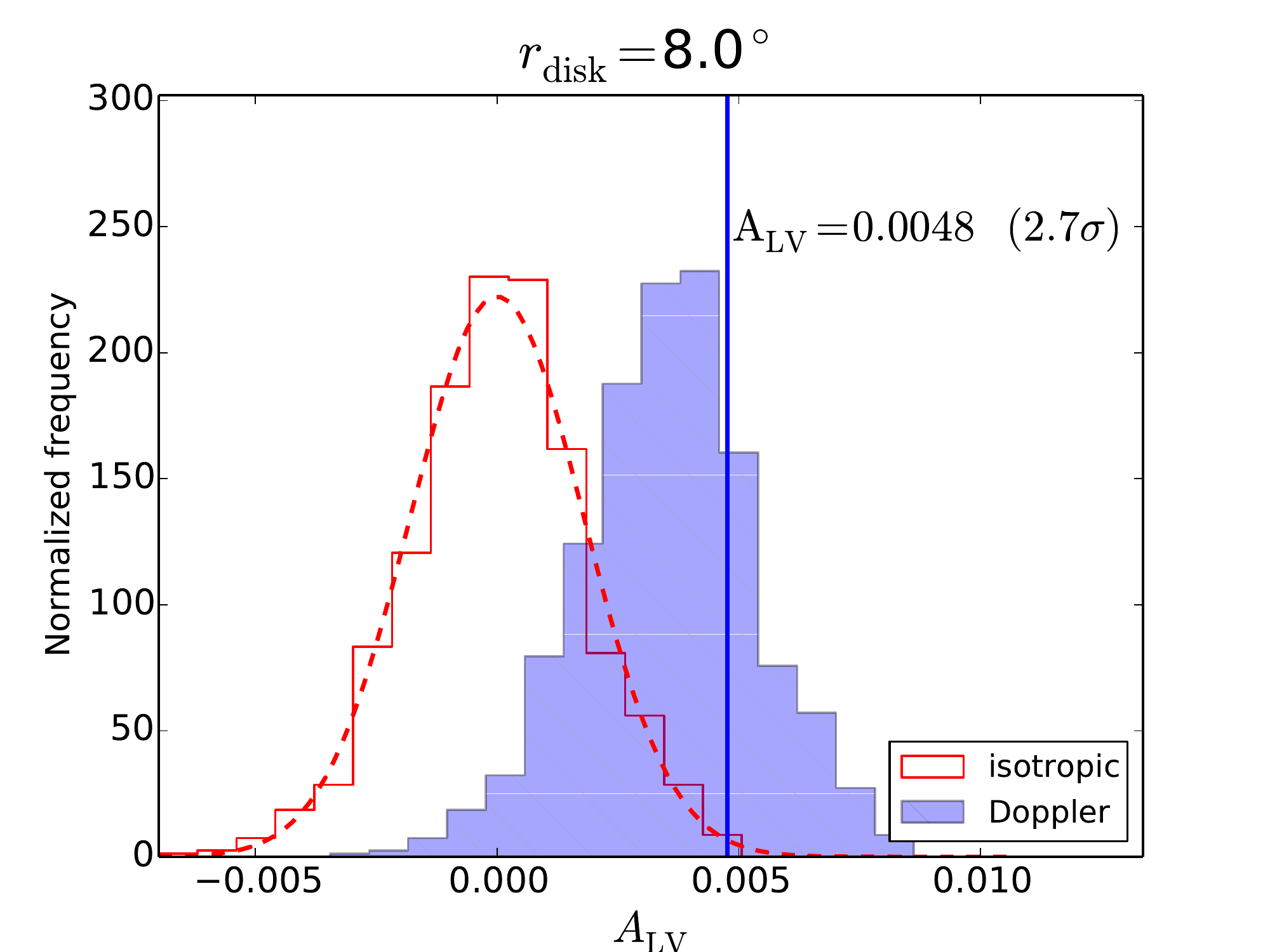}
\caption{The component of local variance dipole amplitudes in the direction $(l,b)=(264^\circ, 48^\circ)$ using disk of radius $\rdisk=0.18^\circ, 8^\circ$ after removing large scales features up to $l_{\rm min}=600$. In both cases, one obtains the Doppler signal at nearly $3\sigma$, while no effect is seen in the orthogonal directions (not shown in figures). The results for our large scale modulation model are not shown above but they mostly coincide with the Doppler models (as we have removed the large scale features), with a very small right shift as expected.}
 \label{fig:doppler18}
\end{figure*}

\begin{figure*}
 \includegraphics[scale=0.65]{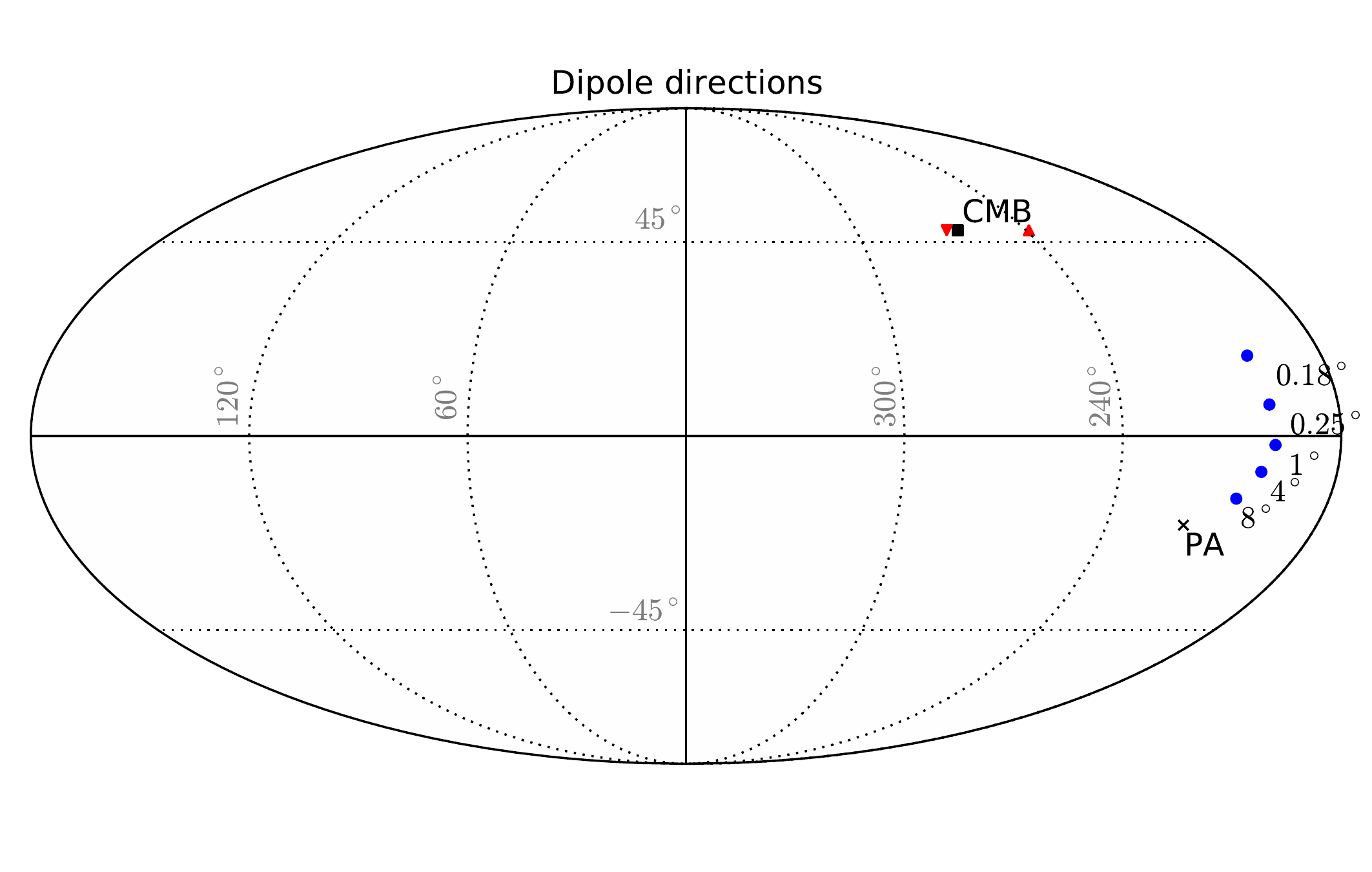}
 \caption{Here we plot some of the directions for the local variance dipoles in our analysis in addition to the CMB dipole direction (black square) and the large scale power asymmetry direction (black cross). The blue dots are directions obtained for the local variance dipole using foreground cleaned 217 GHz Planck maps with the labeled value of $\rdisk$. For smaller $\rdisk$, the dipole direction moves towards the CMB dipole direction. We also plot directions obtained using maps that have large scale features removed (red triangles), with $l_{\rm min}=600\;({\rm triangle\; up})$ and $l_{\rm min}=900\;({\rm triangle\;down})$.}
 \label{fig:dirs}
\end{figure*}

\subsection{The Doppler dipole in local variance maps}
We remove large scale features by simply filtering a map using a high-$l$ filter i.e. set the low-$l$ $a_{lm}$ values to zero in multipole space. Since we are looking for an expected signal that is a vector (i.e. to test the significance of the measurement we need to look at both the direction and magnitude of the dipole signals), we will determine the distribution of the component of the dipoles obtained in the direction of the known CMB dipole. This was also the approach taken by the Planck Collaboration \cite[]{2013arXiv1303.5087P}; the quantity whose distribution they plot, called $\beta_{\parallel}$, is the component of their estimator in the CMB dipole direction.

In Figure \ref{fig:doppler18}, we show results for our local variance analysis using disk radius $\rdisk=0.18^\circ$ but using CMB temperature maps with $a_{lm}=0, $ for $l\leq l_{\rm min}=600$ i.e. large scale features removed up to $l=600$. The signal in the direction parallel to the CMB dipole is consistent with the Doppler model distribution while the isotropic model is disfavored at $3.2\sigma$. When using a larger disk radius $\rdisk=8^\circ$ (also shown in Figure \ref{fig:doppler18}), we measure the Doppler dipole signal at approximately $2.7\sigma$.

We have repeated this analysis using other disk radii, different $l_{\rm min}$ and the 143 GHz channel map. The results are summarized in Table \ref{table:dopplerresults}. Unexpectedly, we obtain higher values of amplitude for the 143 GHz channel but in most cases the dipole amplitude in the direction of CMB dipole is within $2\sigma$ of the expected distribution obtained from the Doppler model. The only exception being the case of $l_{\rm min}=900$, 143 GHz, $\rdisk=0.18^\circ$ in which the signal in data is at $3.1\sigma$ from the Doppler model distribution. In all cases, no amplitude in excess of $2\sigma$ is obtained in two directions orthogonal to the CMB dipole and with each other (see Figure 2 of \cite{2013arXiv1303.5087P}).
\begin{table}
\begin{center}
\begin{tabular}{|c|c|c|c|}
  \cline{3-4}
   \multicolumn{2}{c}{} & \multicolumn{2}{|c|}{Amplitude, $\Alv$ (significance, n$\sigma$)}\\
\hline
   $l_{min}$ & $\rdisk$ & $217$ GHz & $143$ GHz \\
\hline
  600 & 0.18 & 0.0042 $(3.2\sigma)$ & 0.0057 $(3.4\sigma)$ \\
  900 & 0.18 & 0.0027 $(2.9\sigma)$ & 0.0047 $(4.1\sigma)$ \\\hline
  600 & 8.0 & 0.0048 $(2.7\sigma)$ & 0.006 $(2.4\sigma)$ \\
  900 & 8.0 & 0.0042 $(2.6\sigma)$ & 0.0071 $(2.9\sigma)$	\\ \hline
\end{tabular}
\caption{Summary of the Doppler dipole detection results. The significance of detection is computed with respect to the corresponding isotropic distribution for local variance amplitudes (1000 simulations) in the CMB dipole direction.}
\label{table:dopplerresults}
\end{center}
\end{table}

The local variance dipole directions obtained for various cases are shown in Figure \ref{fig:dirs}.

\subsection{Small scale power asymmetry}
We can also investigate the presence of power asymmetry at small scales in the direction $(l,b)=(218^\circ, -20^\circ)$, in our small scale maps that have $a_{lm}$'s up to $l=600$ set to zero. A similar study of the power asymmetry at small scales (but in multipole space) was performed in \cite{2013JCAP...09..033F} using foreground cleaned SMICA maps; they report no such small scale asymmetry after accounting for a number of effects, including estimates of power asymmetry due to the Doppler effect. They report an upper bound on the modulation amplitude of $0.0045\;(95\%)$ at these scales. Another important previous work on constraining hemispherical power asymmetry at smaller scales is \cite{2009JCAP...09..011H} using quasars that reports $-0.0073<\AT<0.012\; (95\%)$ at $k_{\rm eff} \approx 1.5 h {\rm Mpc^{-1}}$.

\begin{figure*}
\centering
 \includegraphics[scale=0.37]{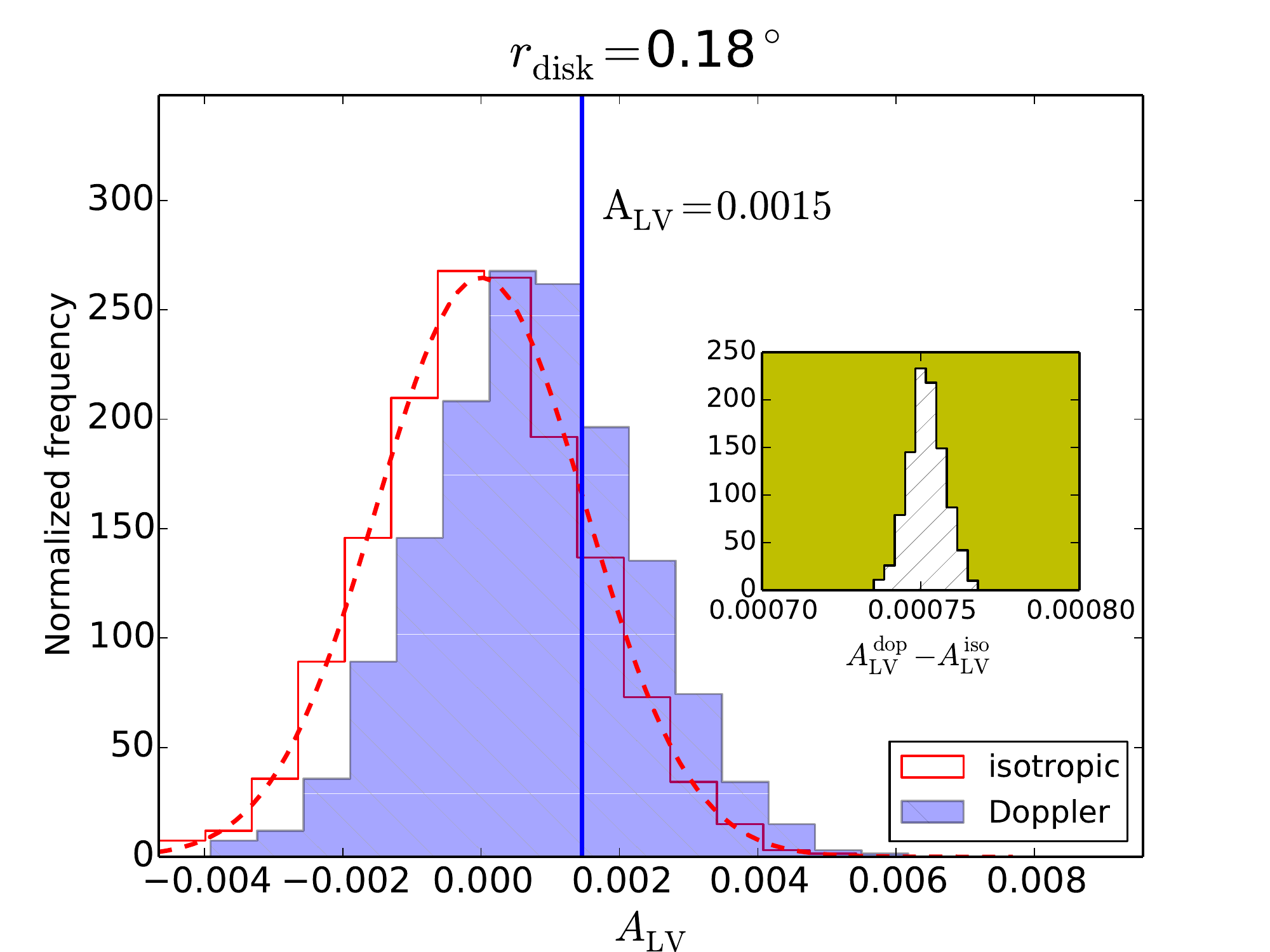}
 \includegraphics[scale=0.37]{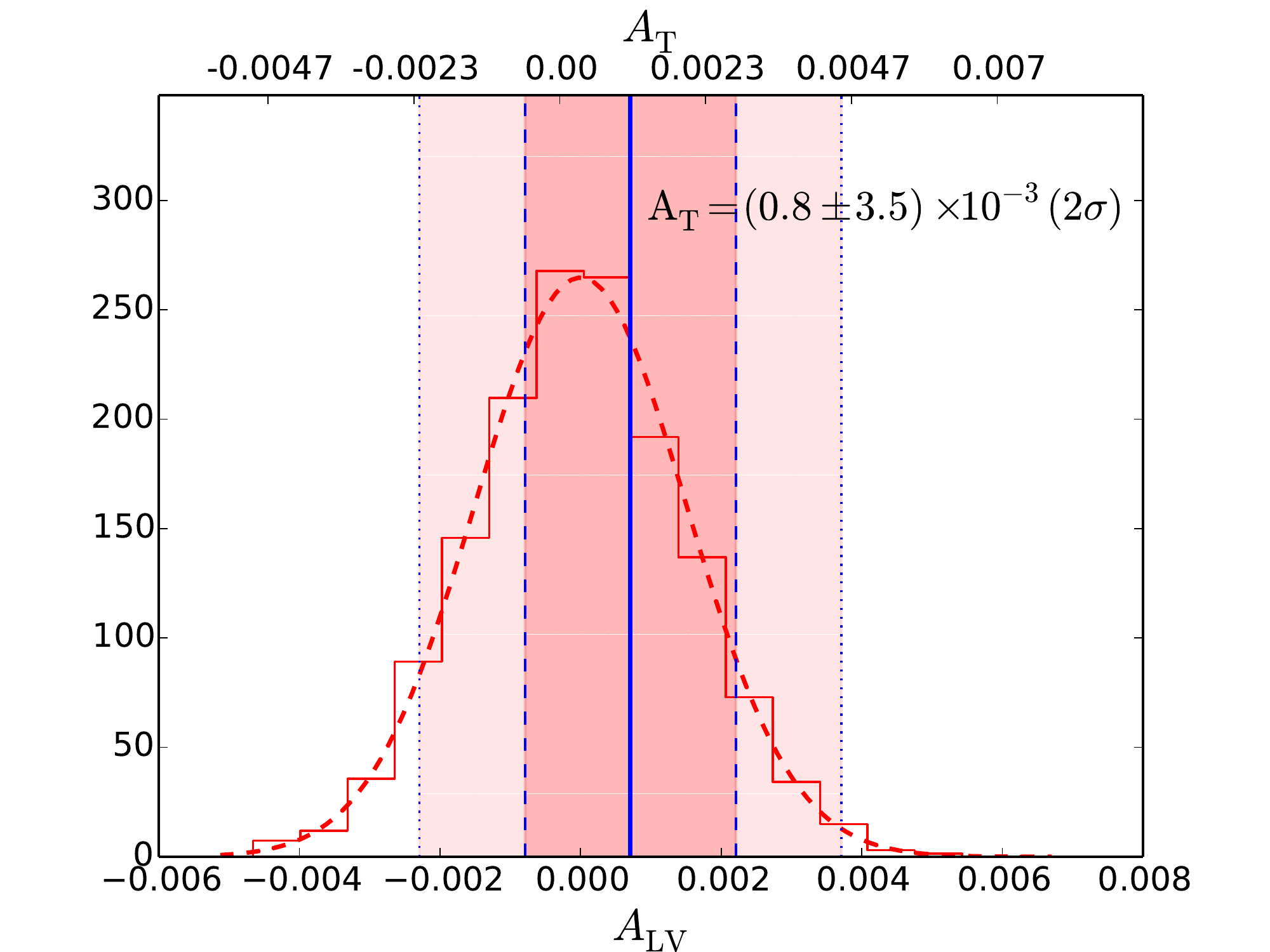}
 \caption{Left: Small scale power asymmetry in the direction $(l,b)=(218^\circ, -20^\circ)$ before subtracting the contribution from the Doppler dipole. The inset shows the distriution for the difference between local variance dipole amplitudes for the Doppler model and the isotropic model. This value $0.00075\pm0.00001 (2\sigma)$ is subtracted from the power asymmetry in data to obtain the constraint for the intrinsic power asymmetry. Right: Small scale power asymmetry in the direction $(l,b)=(218^\circ, -20^\circ)$ after subtracting the Doppler contribution. We also show the corresponding $\AT$ on the top horizontal axis, in addition to $A_{\rm LV}$ on the bottom axis; the fact that these are different is discussed in the text. The error bands show the expected $1\sigma$ and $2\sigma$ errors based on the variance of the isotropic distribution.}
 \label{fig:smallscale}
\end{figure*}

When repeating our measurements shown in Figure \ref{fig:doppler18}, but now in the direction $(l,b)=(218^\circ, -20^\circ)$, we find that the local variance dipole amplitude component in the data is within $1\sigma$ expectation from both isotropic and Doppler model distributions. This is shown in Figure \ref{fig:smallscale}. Each value in the Doppler model distribution is basically shifted right from the isotropic distribution (see the inset in Figure \ref{fig:smallscale}); this shift is proportional to the cosine of the angle between the direction of large scale asymmetry and the direction of the CMB Doppler dipole. Therefore, we subtract this shift to get the value of the intrinsic local variance power asymmetry. Using $\rdisk=0.18^\circ$, we obtain (at $2\sigma$):
\begin{eqnarray}
  \Alv=(0.71\pm3.0)\times 10^{-3}
  \label{eqn:smallconstraint}
\end{eqnarray}

However, we can see from our results for the Doppler model distributions that the relation between $\AT$ of a dipole modulated model and the most likely value of $\Alv$ obtained from the corresponding local variance maps is not quite simple. This can be directly seen in Figure \ref{fig:doppler18} in which the Doppler model had $\AT=3.07\times 0.00123\approx0.0038$ whereas the obtained distributions for $\Alv$ are different for the two cases with different $\rdisk$. To estimate the $\AT$ constraint from the $\Alv$ constraint in eqn \ref{eqn:smallconstraint}, let us assume that the relationship between them is linear for small values of $\AT$ up to approximately $\AT=0.0038$ (for a given frequency channel, $\rdisk$ and $l_{\rm min}$). Then, for the case of the 217 GHz channel maps, $\rdisk=0.18^\circ$ and $l_{\rm min}=600$, we can use the correspondence between $\AT$ and $\Alv$ of the distributions from simulations in Figure \ref{fig:doppler18}, and translate the constraint in eqn \ref{eqn:smallconstraint} to (at $2\sigma$):
\begin{eqnarray}
 \AT=(0.8\pm3.5)\times 10^{-3}
\end{eqnarray}
To further check our translation between $\Alv$ and $\AT$, we repeated this small scale analysis with a $A_{\rm T, \textit{input}}=0.0008$ modulation model and recovered the intrinsic local variance dipole amplitude obtained in eqn \ref{eqn:smallconstraint}. Using other larger values of $\rdisk$ or the 143 GHz channel map, we obtained slightly weaker but consistent constraints for the small scale power asymmetry $\AT$ in the direction of the large scale hemispherical power asymmetry.

\section{Discussion and Summary}
\label{sec:discussion}
In this short work, we have used local variance maps to study the power asymmetry in Planck temperature anisotropy maps, extending the work done in \cite{2014ApJ...784L..42A} that used the same method to smaller scales. We have shown that the effect of the Doppler dipole is small for local variance dipole measurements at large disk radii ($\rdisk\gtrsim 4^\circ$); we find that the peak of the Doppler model distribution for the 217 GHz case is less than $0.5\sigma$ away from that of the isotropic distribution for the $\rdisk=4^\circ$ result. The difference gets even smaller for larger values of $\rdisk$. Further, we have also shown that the Doppler dipole can be measured to a moderate significance using this method.  The first measurement of the Doppler signal in the temperature fluctuations was done by the Planck Collaboration in \cite{2013arXiv1303.5087P}. The method of local variance is a pixel space method, so our measurement is complimentary to that done by the Planck team using harmonic space estimators sensitive to the correlations between different multipoles induced by the Doppler effect. However, local variance dipoles are not sensitive to the aberration effect and therefore our method is less sensitive than the estimators that consider the aberration effect in addition to the dipolar power modulation. 

For our large scale asymmetry analysis, we find similar results using both 217 GHz channel maps (reported in figures in this paper) and 143 GHz channel maps. This is in agreement with previous works \cite[]{2009ApJ...699..985H, Hansen:2004vq} that have investigated channel dependence of hemispherical power asymmetry in WMAP maps. 

Throughout our analysis, we included a large scale modulation model in which temperature anisotropies were modulated only for scales smoothed (Gaussian smoothing) at fwhm=1.25 degrees and found that the anomalous power asymmetry seen at all $\rdisk$ analyzed in this paper using the local variance method is consistent with this simple phenomenological model. This  tells us about the scale dependence of the power asymmetry seen in data as our modulation model generates a scale dependent hemispherical asymmetry in $C_l$s (see Figure \ref{fig:modulation}).

We summarize important aspects of our work and results in the following points:
\begin{itemize}
 \item We have verified the hemispherical power asymmetry results obtained in \cite{2014ApJ...784L..42A}, with the exception of $\rdisk=1,2$ degrees for which we have identified the need to use more disks in order to cover the whole sky.
 \item Once we use smaller disk radii in our local variance analysis, the effect of the dipolar Doppler modulation becomes increasingly important which can be directly observed from our dipole amplitude distributions.
 \item After removing large scale features up to $l=l_{\rm min}=\{600,900\}$, we could detect the expected dipolar Doppler modulation in Planck temperature anisotropy maps at a significance of approximately $3\sigma$.
 \item We have obtained a constraint on dipolar modulation amplitude at small scales ($l>600$) in the direction of the large scale anomalous hemispherical power asymmetry as $\AT=0.0008\pm0.0035 (2\sigma)$. 
\end{itemize}

We expect this work to be useful for a better understanding of the power asymmetries that are known to exist in the CMB data. In particular, we hope that our work will shed more light on local variance statistics in CMB maps which is already being used to compare theoretical models of power asymmetry with data \cite[]{2014arXiv1408.3057J}. While a satisfying theoretical explanation for the power asymmetry anomaly still lacks in the literature, several interesting proposals exist \cite[]{2013PhRvD..87l3005D, 2013PhRvL.110a1301S, 2013JCAP...08..007L, 2008PhRvD..78l3520E, 2009PhRvD..80h3507E, 2013PhRvD..88h3527N, 2014MNRAS.442..670C}. The statistical fluke hypothesis still remains a possibility too \cite[]{2011ApJS..192...17B}. More careful studies of both data and theoretical possibilities are important since the implications of progress in any direction consists of learning more about the fundamental statistical assumptions that we make about the universe at large scales. 

\section*{Acknowledgments} This work is supported by the National Aeronautics and Space Administration under Grant No. NNX12AC99G issued through the Astrophysics Theory Program. The computations for the project were performed using computing resources at the Penn State Research Computing and Cyberinfrastructure (RCC). The author would like to thank Yashar Akrami and Sarah Shandera for discussions and many useful suggestions, Donghui Jeong for useful suggestions on an earlier draft of this work, and Julian Borrill for help with getting the Planck \ffp simulation data. In addition, the author would like to thank Shaun Hotchkiss for a critical review and suggesting improvements to the manuscript.

\bibliographystyle{mn2e_eprint}

\label{lastpage}
\end{document}